\documentclass[submission]{eptcs}
\providecommand{\event}{HaTT'16} 
\usepackage[utf8x]{inputenc}
\usepackage[T1]{fontenc}
\usepackage[american]{babel}
\usepackage{graphicx}
\usepackage{xspace}
\usepackage{listings}
\usepackage{stmaryrd}
\usepackage{todonotes}
\usepackage{caption}
\usepackage{subcaption}

\newcommand{\smtcoq}{SMTCoq\xspace}
\newcommand{\coq}{Coq\xspace}
\newcommand{\zchaff}{ZChaff\xspace}
\newcommand{\verit}{veriT\xspace}
\newcommand{\cvcfour}{CVC4\xspace}
\newcommand{\lfsc}{LFSC\xspace}
\newcommand{\ocaml}{OCaml\xspace}

\title{Extending \smtcoq, a Certified Checker for SMT\\(Extended Abstract)}

\author{
Burak Ekici
\institute{The University of Iowa\\Iowa City, USA}
\email{burak-ekici@uiowa.edu}
\and
Guy Katz
\institute{New York University\\New York, USA}
\email{guy.katz@nyu.edu}
\and
Chantal Keller
\institute{LRI, Universit\'e Paris-Sud\\Orsay, France}
\email{Chantal.Keller@lri.fr}
\and
Alain Mebsout
\institute{The University of Iowa\\Iowa City, USA}
\email{alain-mebsout@uiowa.edu}
\and
Andrew J. Reynolds
\institute{The University of Iowa\\Iowa City, USA}
\email{andrew-reynolds@uiowa.edu}
\and
Cesare Tinelli
\institute{The University of Iowa\\Iowa City, USA}
\email{cesare-tinelli@uiowa.edu}
}

\def\titlerunning{\smtcoq}
\def\authorrunning{B. Ekici, G. Katz, C. Keller, A. Mebsout,
  A. J. Reynolds \& C. Tinelli}

\lstdefinelanguage{Coq}%
  {morekeywords={Variable,Inductive,CoInductive,Fixpoint,CoFixpoint,%
      Definition,Program, Lemma,Theorem,Corollary,Axiom,Local,Save,Grammar,Syntax,Intro,%
      Trivial,Qed,Intros,Symmetry,Simpl,Rewrite,Apply,Elim,Assumption,%
      Left,Cut,Case,Auto,Unfold,Exact,Right,Hypothesis,Pattern,Destruct,%
      Constructor,Defined,Fix,Record,Proof,Induction,Hints,Exists,let,in,%
      Parameter,Split,Red,Reflexivity,Transitivity,if,then,else,Opaque,%
      Transparent,Inversion,Absurd,Generalize,Mutual,Cases,of,end,Analyze,%
      AutoRewrite,Functional,Scheme,params,Refine,using,Discriminate,Try,%
      Require,Load,Import,Scope,Prop,Set,Type,Open,Section,End,match,with,Ltac,fun,forall,exists,Canonical,Structure,Eval,Notation,as,return,Goal%
        },%
   sensitive, %
   morecomment=[n]{(*}{*)},%
   morestring=[d]",%
   literate={=>}{{$\Rightarrow$}}1 {>->}{{$\rightarrowtail$}}2{->}{{$\,\,\to\,\,$}}1
   {\/\\}{{$\wedge$}}1
   {|-}{{$\vdash$}}1
   {\\\/}{{$\vee$}}1
   {~}{{$\sim$}}1
   {'}{'}1
   {⟦}{{$\llbracket$}}1
   {⟧}{{$\rrbracket$}}1
   {-->}{{$\longrightarrow$}}1
  }[keywords,comments,strings]%
\begin{document}

\maketitle

\begin{abstract}
  This extended abstract reports on current progress of \smtcoq, a
  communication tool between the \coq proof assistant and external SAT
  and SMT solvers. Based on a checker for generic first-order
  certificates implemented and proved correct in \coq, \smtcoq offers
  facilities both to check external SAT and SMT answers and to improve
  \coq's automation using such solvers, in a safe way. Currently
  supporting the SAT solver \zchaff, and the SMT solver \verit for the
  combination of the theories of congruence closure and linear integer
  arithmetic, \smtcoq is meant to be extendable with a reasonable amount
  of effort: we present work in progress to support the SMT solver
  \cvcfour and the theory of bit vectors.
\end{abstract}



\section{Introduction}

\smtcoq\footnote{\smtcoq is distributed as free software at
  \url{https://github.com/smtcoq/smtcoq}.}~\cite{DBLP:conf/cpp/ArmandFGKTW11}
is a tool that allows the \coq~\cite{barras:cpa} proof assistant to
communicate with external automatic solvers for Boolean
satisfiability (SAT) and Satisfiability Modulo Theories (SMT). 
Its twofold goal is to:
\begin{itemize}
\item increase the confidence in SAT and SMT solvers: \smtcoq
  provides an independent and certified checker for SAT and SMT proof
  witnesses;
\item safely increase the level of automation of \coq: \smtcoq provides
  starting safe tactics to solve a class of \coq goals automatically by
  calling external solvers and checking their answers (following a {\em
    skeptical} approach).
\end{itemize}
\smtcoq currently supports the SAT solver \zchaff~\cite{DBLP:conf/sat/MahajanFM04} and the
SMT solver \verit~\cite{DBLP:conf/cade/BoutonODF09} for the quantifier-free
fragment of the combined theory of linear integer arithmetic and equality with uninterpreted functions.
For this combined theory, \smtcoq's certificate checker has proved
to be as efficient as state-of-the-art certified
checkers~\cite{DBLP:conf/cpp/ArmandFGKTW11,DBLP:conf/itp/BohmeW10}.

There is a large variety of SAT and SMT solvers, with each solver typically
excelling at solving problems in some specific class of propositional 
or first-order problems. While the SAT and SMT communities have adopted
standard languages for expressing {\em input} problems
(namely the DIMACS standard for SAT and the SMT-LIB~\cite{BarST-SMT-10} standard for SMT), 
agreeing on a common {\em output} language for proof witnesses has proven 
to be more challenging. 
Several formats~\cite{DBLP:journals/amai/Gelder12,DBLP:journals/fmsd/StumpORHT13,BessonFT11}
have been proposed but none has emerged as a standard yet.
Each proof-producing solver currently implements its own variant of these formats.

To be able to combine the advantages of multiple SAT and SMT
solvers despite the lack of common standards for representing proof certicates, 
\smtcoq has been designed to be modular along two dimensions:
\begin{itemize}
\item supporting new theories: \smtcoq's main checker is an extendable
  combination of independent {\em small checkers};
\item supporting new solvers: \smtcoq's kernel relies on a generic
  certificate format that can encode most SAT and SMT reasonings for
  supported theories; the encoding can be done during a {\em preprocessing} phase,
  which does not need to be certified.
\end{itemize}

In this abstract, we emphasize the key ideas behind the modularity of
\smtcoq, and validate this by reporting on work in progress on
the integration of the SMT solver
\cvcfour~\cite{DBLP:conf/cav/BarrettCDHJKRT11} and the theory of bit
vectors. We simultaneously aim at:
\begin{itemize}
\item offering to \cvcfour users the possibility to formally check its
  answers in a trusted environment like Coq;
\item bringing the power of a versatile and widely used SMT solver like \cvcfour to
  \coq;
\item providing in \coq a decision procedure for bit vectors, a theory
  widely used, for instance, for verifying circuits or programs using machine
  integers.
\end{itemize}



\section{The \smtcoq Tool}
\label{sec:smtcoq}
\subsection{General Idea}
\label{sec:smtcoq:general}

\begin{figure}[t]
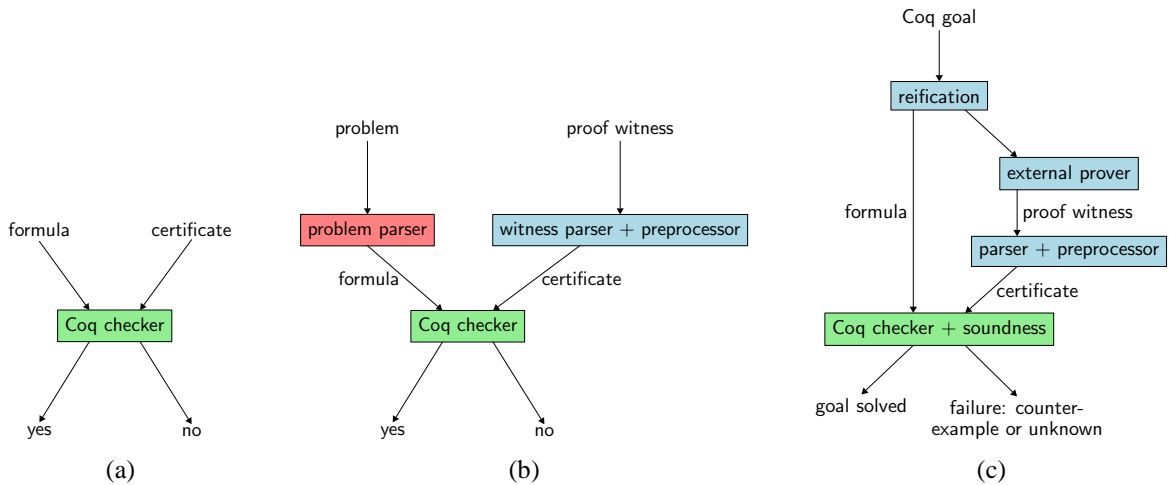

  \centering
    \begin{subfigure}[b]{3cm}
     \includegraphics[width=\textwidth]{checker.mps}
     \caption{}
     \label{fig:check1}
    \end{subfigure}
    \qquad
    \begin{subfigure}[b]{6cm}
     \includegraphics[width=\textwidth]{vernac.mps}
     \caption{}
     \label{fig:check2}
    \end{subfigure}
    \qquad
    \begin{subfigure}[b]{4.7cm}
     \includegraphics[width=\textwidth]{tactic.mps}
     \caption{}
     \label{fig:check3}
    \end{subfigure}
  \caption{\smtcoq's main checker and its uses}
  \label{fig:checker}
\end{figure}

The heart of \smtcoq is a checker for a generic format of certificates
(close to the format proposed by Besson {\em et al.}~\cite{BessonFT11}),
implemented and proved correct inside \coq (see
Figure~\ref{fig:check1}). Taking advantage of \coq's
computational capabilities the SMTCoq checker is fully executable, either
inside \coq or after extraction to a general-purpose
language~\cite{DBLP:conf/types/Letouzey02}.


The \coq signature of this checker is the following:
\begin{lstlisting}
checker : formula ->$\,$ certificate ->$\,$ bool
\end{lstlisting}
where the type \lstinline!formula! represents the deep embedding in \coq
of SMT formulas, and the type\break \lstinline!certificate! represents
\smtcoq's format of certificates.

The checker's soundness is stated with respect to a translation function from the
deep embedding of SMT formulas into \coq terms:
\begin{lstlisting}
$\llbracket\bullet\rrbracket$ : formula ->$\,$ bool
\end{lstlisting}
that interprets every SMT formula into its \coq Boolean counterpart. The
correctness of the checker:
\begin{lstlisting}
checker_sound : $\forall$ f c, checker f c = true -> $\ \llbracket$f$\rrbracket$
\end{lstlisting}
thus means that, given a formula and a certificate for which the checker
answers positively, then the interpretation in \coq of the formula is
valid.

The choice of the type of Booleans \lstinline!bool! as the codomain of
the translation function
\lstinline!$\llbracket\bullet\rrbracket$!, instead of the type of
(intuitionistic) propositions \lstinline!Prop!, allows us to handle the
checking of the classical reasoning made by SMT solvers without adding
any axioms. The SSReflect~\cite{DBLP:journals/jfrea/GonthierM10} plugin for
\coq can be used to bridge the gap between propositions and Booleans for the
theories considered by \smtcoq. The major shortcoming of this approach is that 
it does not allow quantifiers inside goals sent to SMT solvers, although it
does not prevent one from feeding these solvers universally quantified
lemmas. To increase the expressivity of \smtcoq with respect to
quantifiers, one will need to switch to propositions, and handle
classical logic either by axioms or by restricting attention to decidable atoms
of the considered combined theory.

The first use case of this correct-by-construction checker is to check the
validity of a proof witness, or proof \emph{certificate} coming from an external solver against some
input problem (Figure~\ref{fig:check2}). In this use case, the
trusted base is both \coq and the parser of the input problem.
The parse is part of the trusted based because we need to make sure 
we are effectively verifying a proof of the problem we sent to the external solver.
However, this parser is fairly straightforward.

The second use case is within a \coq tactic (Figure~\ref{fig:check3}).
We can give a \coq goal to an external solver and get a
proof certificate for it. If the checker can validate the certificate, 
the soundness of the checker allow us to establish a proof of the initial goal. 
This process is known as {\em computational reflection} as it uses a computation
(here, the execution of the checker) inside a proof. In this use case,
the trusted base consists only of \coq: if something else goes wrong (e.g., the
checker cannot validate the certificate), the tactic will fail, but
nothing unsound will be added to the system.

In both cases, a crucial aspect for modularity purposes is the possibility to
{\em preprocess} proof certificates before sending them to the \smtcoq
checker, 
  without having to prove anything about this preprocessing
  stage
. Again, if the preprocessor is buggy, the checker will fail to
validate the proof certificate (by returning \lstinline!false!), which means
that while nothing is learned, nothing unsafe is added to Coq's context. 
This allows us to easily extend \smtcoq with new solvers: as long as the certificate
coming from the new solver can be logically encoded into \smtcoq's
certificate format, we can implement this encoding at the preprocessing
stage. As a result, \smtcoq's current support for both \zchaff and \verit
is provided through the implementation of a preprocessor for each solver.
Both preprocessors convert to the same proof format, 
thus sharing the same checker.

Using a preprocessor is also beneficial for efficiency: proof certificates
may be encoded more compactly before being sent to the \smtcoq checker, 
which may improve performance.

\subsection{The Checker}
\label{sec:smtcoq:checker}

\begin{figure}[h!]
  \begin{center}
    \includegraphics[height=8cm]{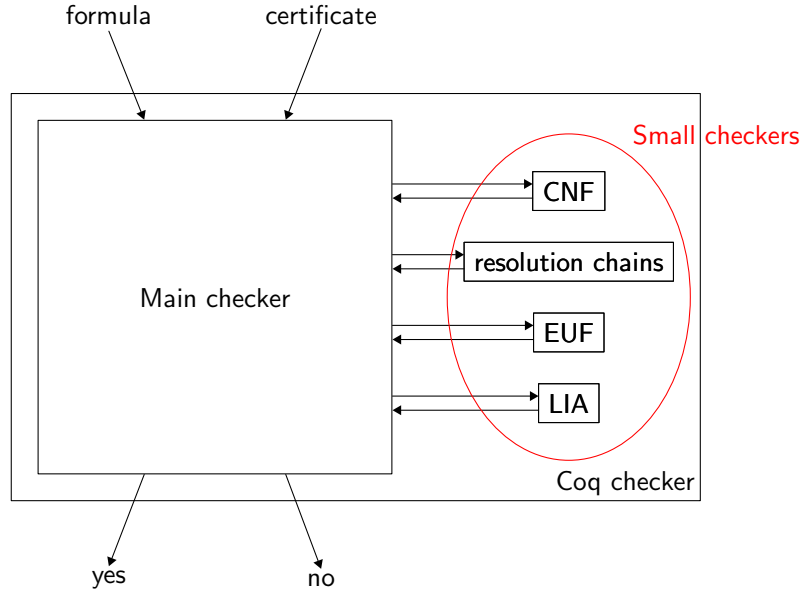}
  \end{center}
  \caption{Internals of the \coq checker}
  \label{fig:schecker}
\end{figure}

We now provide more details on the checker of \smtcoq. As presented in
Figure~\ref{fig:schecker}, it consists of a {\em main checker} obtained as
the combination of several {\em small checkers}, each specialized in one aspect 
of proof checking in SMT (e.g., CNF conversion, propositional reasoning, 
reasoning in the theory of equality, linear arithmetic reasoning, and so on).

The type \lstinline!certificate! is actually the aggregation of
specialized types, one for each small checker. The role of the main checker 
is thus to dispatch each piece of the certificate to its
dedicated small checker, until the initial formula is proved.

A small checker is a \coq program that, given a (possibly empty) list of
formulas and a certificate associated with it (which may be just a piece of
the input certificate), computes a new
formula:
\begin{lstlisting}
small_checker : list formula -> certificate_sc -> formula
\end{lstlisting}
The soundness of the \lstinline!checker! comes from the soundness of
each small checker, stated as follows:
\begin{lstlisting}
small_checker_sound : $\forall$ f$_1$ $\dots$ f$_n$ c,
    $\llbracket$f$_1$$\rrbracket$ $\wedge$ $\dots$ $\wedge$ $\llbracket$f$_n$$\rrbracket$ -> $\ \llbracket$small_checker [f$_1$;...;f$_n$] c$\rrbracket$
\end{lstlisting}
meaning that the small checker returns a formula which is implied (after
translation into \coq's logic) by the conjunction of its premises. Note
that the list of premises may be empty: in such a case,
the small checker returns a tautology in Coq.

Here are some examples of small checkers.
\begin{itemize}
\item For propositional resolution chains, the checker takes as input a list of 
  premises and returns a resolvent if it
  exists, or a trivially true clause otherwise. In this case, a
  certificate is not required as part of the small checker's input.
\item For the theory of equality with uninterpreted functions (EUF),
the checker takes as input a formula in this theory formulated as a certificate
  (corresponding to a theory lemma produced by the SMT solver), and
  returns the formula if it is able to check it, or a trivially true
  clause otherwise. In this case, no premises are given.
\item For linear integer arithmetic (LIA), the checker works similarly to the
  EUF checker, but checks the formula using Micromega~\cite{DBLP:conf/types/Besson06}, an
  efficient decision procedure for this theory implemented in \coq.
\end{itemize}

The only thing that small checkers need to share is the type
\lstinline!formula!, and its interpretation into \coq Booleans. Each
small checker may then reason independently, using separate pieces of
the certificate. Again, this is crucial for modularity: to extend \smtcoq
with a new theory, one only has to extend the type \lstinline!formula!
with the signature of this theory and, 
  independently of the already
  existing checkers
, implement a small checker for this theory and
prove its soundness.

Notice that ``small checker'' can be understood in a very general sense: any 
function that, given a list of first-order formulas, returns an implied
first-order formula, can be plugged into SMTCoq as a small checker. 
In principle, such a checker could even be as complex as an SMT solver,
as long as it can be proved correct in Coq.



\section{Work in Progress: Extensions to \cvcfour and Bit Vector
  Arithmetic}
\subsection{Support for \cvcfour}

\cvcfour is a proof-producing SMT solver, whose proof format is based on
the Logical Framework with Side
Conditions (\lfsc)~\cite{DBLP:journals/fmsd/StumpORHT13}. \lfsc extends 
the Edinburgh Logical Framework (LF)~\cite{DBLP:journals/jacm/HarperHP93} by allowing types
with computational \emph{side conditions}, explicit computational checks 
defined as programs in a small but expressive functional
first-order programming language. 
The language has built-in types for arbitrary precision integers and rationals, 
ML-style pattern matching over \lfsc type constructors, recursion, 
a minimal support for exceptions, and 
a very restricted set of imperative features.
One can define proof rules in LFSC as typing rules that may optionally include 
a side condition written in this language.
When checking the application of such proof rules, an \lfsc checker
computes actual parameters for the side condition and executes its code;
if the side condition fails, the \lfsc checker rejects the rule application.
The validity of an \lfsc proof witness thus relies on the correctness of 
the side condition functions used in the proof.
\lfsc comes with a set of pre-defined side conditions for various
theories, used by the \cvcfour proof production mechanism.

The key differences between \lfsc and the \smtcoq format are presented
in Table~\ref{tab:lsfc-vs-smtcoq}.

\begin{table}[h!]
  \begin{center}
    \begin{tabular}{|r|c|c|}
      \hline
       & \lfsc & \smtcoq\\
      \hline
      Rules & deduction + computation & deduction + certificate\\
      \hline
      Nested proofs & supported & not supported\\
      \hline
    \end{tabular}
  \end{center}
  \caption{Main differences between the \lfsc and \smtcoq certificate formats}
  \label{tab:lsfc-vs-smtcoq}
\end{table}

The major difference lies in the presentation of the deduction rules. In
\smtcoq, the small checkers deduce a new formula from
already known formulas, possibly with the help of a piece of certificate
that depends on the theory. The \lfsc format is more uniform, thanks to
the side conditions described above.

To support \lfsc, and so \cvcfour, we are in the process of implementing (in
\ocaml) an untrusted preprocessor that transforms \lfsc proofs into \smtcoq
proofs. To this end, for some theories, we need to replay parts of the
side conditions, in order to produce the corresponding \smtcoq premises,
conclusion and piece of certificate that will be passed to the small
checkers. This encoding, however, is relatively straightforward:
\begin{itemize}
\item for propositional reasoning, \lfsc side conditions use the same logical content as \smtcoq
  rules;
\item CNF conversion and EUF proofs are nested in LFSC, so they require some
  processing for the moment;
\item for linear integer arithmetic, since \smtcoq relies on an existing
  decision procedure in \coq, it only needs to know what theory lemma is being proved, 
  and can ignore the actual proof steps in the \lfsc certificate.
\end{itemize}

One difficulty in translating \lfsc proofs to the \smtcoq format
comes from to the possibility in \lfsc of using natural-deduction-style proofs, 
where one can nest one proof inside another. 
For instance, it is possible to have lemmas inside an \lfsc proof 
whose witnesses are themselves \lfsc proofs. 
The architecture of the main and small checkers of \smtcoq does not currently allow 
this sort of nesting:
every clause produced by the small checkers needs to be a direct consequence of
input clauses or clauses that were previously produced. To encode an
\lfsc proof into \smtcoq, our preprocessor thus linearizes nested
proofs. The \lfsc proofs generated by \cvcfour are constructed in such a
way that this does not cause a blow-up in practice;
however, to support \lfsc in general, we plan to extend \smtcoq
certificates with nested proofs. Again, this extension should be made
easier by the modularity inside the checker. It should impact only the main
checker, and not the various small checkers already in \smtcoq.

\subsection{Support for Bit Vector Arithmetic}

\cvcfour has been recently extended to produce LFSC proofs for the quantifier-free
fragment of the SMT theory of bit vectors~\cite{DBLP:conf/lpar/HadareanBRTD15}. 
To check proof certificates in this theory, \smtcoq needs be extended with it. 
As explained in Section~\ref{sec:smtcoq:checker}, to do that one needs to:
\begin{enumerate}
\item\label{it:form} extend the \coq representation of formulas with the
  signature of the bit vector theory and the interpretation function
  into \coq terms;
\item\label{it:checker} implement (new) small checkers and their corresponding
  certificates for this theory, 
  and prove their correctness.
\end{enumerate}

Step~\ref{it:form} is a simple extension on the \smtcoq side. The major
difficulty is that \coq itself has limited support for bit vectors. Its
bit vector library provides only the implementation of bitwise
operations (and not arithmetic operations), and no proofs. We are thus
currently implementing a more complete library for this theory.
Step~\ref{it:checker} involves implementing and adding new certified \coq programs 
(the small checkers).
As mentioned, however, because of \smtcoq's design, 
none of the previous small checkers and their proofs of correctness need to be changed
as a result of this addition.

\lfsc proofs for bit vectors produce by \cvcfour mainly involve the following two kinds
of deduction steps:\
\begin{itemize}
\item \emph{bit-blasting} steps that reduce the input bit vector formula 
 to an equisatisfiable propositional formula;
\item standard propositional reasoning steps (based on resolution).
\end{itemize}
The propositional steps can be handled directly by previous small
checkers. For the bit-blasting steps, we implemented new small checkers
that relate terms of the bit vector theory with lists of Boolean
formulas representing their bits; we are currently working on producing proofs 
of correctness in \coq for these small checkers.

\lfsc proofs generated by \cvcfour involve a third kind of
step: formula simplifications based on the equivalence of two
bit-vector terms or atomic formulas (for instance, by normalizing inequalities).
Currently, these simplification steps are not provided a detailed \lfsc subproof
by \cvcfour, although there are plans to do so in the near future.
In the current \smtcoq implementation then, we assume those steps, as in the \lfsc
proof coming from \cvcfour, or let the user prove them, in the case of tactics. 
Since those steps correspond to applications of \cvcfour-defined rewriting and 
simplification rules, we plan for now to prove the correctness of these rules once and 
for all at the \coq level, and to pre-process simplification steps into applications 
of these rules.


\section{Related Work}
In addition to related work already discussed throughout the paper,
we now briefly mention a few more notable projects.
Heule {\em et al.} implemented an efficient checker for state-of-the-art
SAT techniques, verified in
ACL2~\cite{DBLP:conf/cade/HeuleHW13,DBLP:conf/itp/WetzlerHH13}. It is
mainly based on a generalization of extended
resolution~\cite{Tseitin70,DBLP:journals/dam/Kullmann99} and on reverse
unit propagation~\cite{DBLP:journals/amai/Gelder12}. \smtcoq currently
handles only standard extended resolution for its propositional part.

Efficient proof reconstruction for SAT and SMT solvers has been
implemented in proof assistants based on higher-order
logic~\cite{Weber08,DBLP:conf/itp/BohmeW10}. Some of these
reconstructions also handle the theory of bit
vectors~\cite{DBLP:conf/cpp/BohmeFSW11}. This approach is based on
translating SAT/SMT certificates to applications of the inference
rules of the kernels of these proof assistants. In contrast, our approach
in \coq is based on computational reflection: the certificate is
directly processed by the reduction mechanism of \coq's kernel.

Based on an efficient encoding of a large subset of HOL goals into
first-order logic, the Sledgehammer
tactic~\cite{DBLP:conf/lpar/PaulsonB10} allows HOL-based proof
assistants to efficiently and reliably help manual proving. Proofs are
replayed using either the proof reconstruction mechanism described above
or a built-in first-order prover. We hope that \smtcoq can help in adding
such techniques into \coq and other Type Theory-based proof assistants,
by providing a proof replay mechanism based on certificates.



\section{Conclusion and Future Work}

\smtcoq has been designed to be modular in such a way that facilities
its extension
with new solvers and new theories. In particular, such extensions
should not require any changes in existing checkers or in their proofs of soundness.
Thus, while it may 
require some effort to certify new small checkers or to translate new
proof formats into the
\smtcoq format, such extensions require only local changes.
 Our current extensions to \cvcfour and bit vectors
arithmetic validate this goal: indeed, the work so far
consisted mostly in implementing an untrusted preprocessor for
certificates and adding new, 
independent checkers. One limiting aspect of \smtcoq is the lack of support for
nested proofs, which we plan to add. Thanks to the modularity of
the checker, we believe this feature too can be added locally.

In the future we plan to continue extending the expressivity of
\smtcoq, and in particular
to offer support for the SMT theory of arrays (for which \cvcfour is
also proof-producing). We believe we can match, and perhaps even
improve upon existing work in terms of efficiency.

The current major limitation of \smtcoq resides in its set of tactics:
presently, it can only handle goals that are directly provable by SMT
solvers, without much encoding of \coq logic into first-order logic. Our
longer term plan is to combine ongoing work on {\em
  hammering}~\cite{DBLP:journals/jfrea/BlanchetteKPU16} for proof
assistants based on Type Theory (such as \coq) with the certificate
checking capabilities offered by \smtcoq.

\paragraph{Acknowledgments.}
We wish to thank the anonymous reviewers for their helpful and constructive
feedback.
This work was supported in part by the Air Force Research Laboratory (AFRL) and the Defense Advanced Research Projects Agency (DARPA) under contracts FA8750-13-2-0241 and FA8750-15-C-0113. Any opinions, findings, and conclusions or recommendations expressed above are those of the authors and do not necessarily reflect the views of AFRL or DARPA.


\bibliographystyle{eptcs}
\bibliography{biblio}
\end{document}